\documentclass[sigconf]{acmart}
\usepackage{multirow}
\usepackage{algorithm}
\usepackage{algorithmic}
\usepackage{cleveref}

\newtheorem{prop}{Proposition}

\copyrightyear{2022}
\acmYear{2022}
\setcopyright{rightsretained}
\acmConference[DAC '22]{Proceedings of the 59th ACM/IEEE Design Automation Conference (DAC)}{July 10--14, 2022}{San Francisco, CA, USA}
\acmBooktitle{Proceedings of the 59th ACM/IEEE Design Automation Conference (DAC) (DAC '22), July 10--14, 2022, San Francisco, CA, USA}
\acmDOI{10.1145/3489517.3530511}
\acmISBN{978-1-4503-9142-9/22/07}

\begin{document}
	
	\title{Pursuing More Effective Graph Spectral Sparsifiers via Approximate Trace Reduction}
	
	\author{Zhiqiang Liu, ~ Wenjian Yu}\authornote{This work is supported by National Key R\&D Program of
China (2019YFB2205002) and NSFC under grant No. 62090025.}
	\affiliation{%
		\institution{Dept. Computer Science \& Tech., BNRist, Tsinghua Univ.}
 		\state{Beijing 100084}
 		\country{China} \\
Email: liu-zq20@mails.tsinghua.edu.cn,   yu-wj@tsinghua.edu.cn
	}
	

	\begin{abstract}
		Spectral graph sparsification aims to find ultra-sparse subgraphs which can preserve spectral properties of original graphs. In this paper, 
		a new spectral criticality metric based on trace reduction is first introduced for identifying spectrally important off-subgraph edges. Then, a physics-inspired truncation strategy and an approach using approximate inverse of Cholesky factor are proposed to compute the approximate trace reduction efficiently. Combining them with the iterative densification scheme in \cite{feng2019grass} and the strategy of excluding spectrally similar off-subgraph edges in \cite{fegrass}, we develop a highly effective graph sparsification algorithm. The proposed method has been validated with  various kinds of graphs. 
		Experimental results show that it always produces sparsifiers with remarkably better quality than  the state-of-the-art GRASS \cite{feng2019grass} in same computational cost, enabling more than 40\% time reduction  for preconditioned iterative equation solver on average. In the applications of power grid transient analysis and spectral graph partitioning, the derived iterative solver  shows 3.3X or more advantages on runtime and memory cost, over the approach based on direct sparse solver.
	\end{abstract}
	
	
	\maketitle
	\section{Introduction}
	Spectral methods, originated from spectral graph theory, are playing increasingly important roles in many real problems, such as on-chip power grid analysis, partial differential equation solution, spectral graph partitioning, and semi-supervised learning, etc \cite{icm10}.
	
	Spectral graph sparsification aims to find ultra-sparse subgraphs (called sparsifiers) which can preserve spectral properties of original graphs. In the past decades, spectral sparsification approaches have been extensively studied in both theory \cite{EffRes08,bss,kmp} and practice \cite{feng16,feng18,feng2019grass,fegrass,sfgrass,pgrass,hypersf}. An effective resistance based sampling method was proposed in \cite{EffRes08}. However, computing effective resistances with respect to general graphs can be extremely time-consuming even with the state-of-the-art method based on Johnson–Lindenstrauss (JL) theorem. Another approach exploiting effective resistances in spanning tree instead of the original graph was proposed in \cite{kmp}, which usually causes a much greater number of edges recovered for achieving similar spectral approximation level. The ``BSS process" proposed in \cite{bss} can construct $\epsilon$-sparsifiers with $O(n\epsilon^{-2})$ edges for every graph, but the cubic time complexity prevents it from being applied to large-scale practical problems.
	
	GRASS proposed in \cite{feng16,feng18,feng2019grass} is the first practically-efficient  spectral graph sparsification algorithm. It leverages spectral perturbation analysis for identifying and recovering spectrally-critical off-tree edges and can produce high-quality spectral sparsifiers (low relative condition number of graph Laplacians). Two different approaches were then proposed in \cite{sfgrass,fegrass} to speed up the graph sparsification phase. SF-GRASS in \cite{sfgrass} leverages spectral graph coarsening and graph signal processing  techniques, while feGRASS in \cite{fegrass} is based on effective edge weights and a concept of spectral edge similarity. The both 
	approaches can largely reduce the runtime of graph sparsification. However, in terms of the approximation quality of the produced sparsifier, GRASS is still the state-of-the-art.
	
	It should be noted that there are many applications, where the graph sparsifier needs to be extracted once and can be reused many times. For example, in transient simulation of power grid, the linear equation systems with the same or similar coefficient matrices are solved and they can share one sparsifier for constructing the preconditioner in iterative equation solution. In these applications, the time for constructing the sparsifier can be amortized and the approximation quality of sparsifier dominates the overall performance. So, more effective graph sparsification algorithms, which can produce sparsifiers with higher approximation quality, are highly demanded.
	
	In this work, we aim to develop a more effective graph spectral sparsification algorithm. Our main contributions are as follows.
	
	1) A metric for the spectral criticality of off-subgraph edge is proposed, which considers the 
trace of matrix $L_S^{-1}L_G$ as a proxy of relative condition number of Laplacians $L_G$ and $L_S$ for optimization.
	
	
	2) A physics-inspired truncation strategy and an approach based on computing approximate sparse inverse of Cholesky factor are proposed to  compute the approximate trace reduction with respect to general subgraphs efficiently.
	
	3) Combining these techniques with the iterative densification scheme in \cite{feng2019grass} and the strategy of excluding spectrally similar off-subgraph edges in \cite{fegrass}, we develop a highly effective graph sparsification algorithm. Extensive experiments have been carried out to validate the effectiveness of the proposed method, which always produces  sparsifiers with remarkably better quality than GRASS \cite{feng2019grass}. And, it derives an efficient iterative equation solver superior to the direct sparse solver \cite{cholmod2008} on runtime and memory usage.
	
	
	\section{Background}
	
	Consider a weighted undirected graph $G =(V, E, w)$, where $V$ and $E$ denote the sets of vertices (nodes) and edges, respectively. $w$ is a positive weight function. We use $w_{i,j}$ to denote the weight of edge $(i,j)$. The Laplacian matrix of $G$ is denoted by $L_G \in \mathbb{R}^{n \times n}$. 
	\begin{equation}
	\setlength{\abovedisplayskip}{3pt}
	\setlength{\belowdisplayskip}{3pt}
	L_G(i,j)=\left\{
	\begin{aligned}
	-w_{i,j}, \quad (i,j) \in E \\
	\sum_{(i,k) \in E} w_{i,k}, \quad i=j \\
	~ ~ 0, \quad \quad \textrm{otherwise} ~ .
	\end{aligned}
	\right.
	\end{equation}
	$L_G $ is singular as the smallest eigenvalue is 0. We assume that some small values are added to diagonal elements  to make the Laplacian matrix invertible. To simplify the notations, we still use $L_G$ to denote the resulted symmetric diagonally dominant (SDD) matrix.  
	
	Graph spectral sparsification aims to find an ultra-sparse subgraph $P$ (called sparsifier) which is spectrally similar to the original graph $G$. The sparsifiers can be utilized to speed up iterative solution of SDD matrices. For example, taking $L_P$ as preconditioner, the preconditioned conjugate gradient (PCG) algorithm can find an $\epsilon-$accurate solution in at most $O(\sqrt{\kappa(L_G,L_P)} \log \frac{1}{\epsilon})$ iterations. $\kappa(L_G,L_P)$ denotes the relative condition number of $L_G$ and $L_P$. 
	
	In existing work, graph spectral sparsification typically involves the following two steps \cite{feng2019grass,feng16,feng18,fegrass,sfgrass}:
	
	1). Extract a spectrally critical spanning tree from $G$;
	
	2). Recover a few spectrally critical off-tree edges from $G$ and add them into the spanning tree to form subgraph $P$.
	
	Recovering too few edges results in poor similarity, whereas too many edges can lead to high computational cost. To obtain high-quality sparsifier needs to identify the most spectrally important off-subgraph edges. So, spectral criticality metrics which reflect the spectral importance of each edge are desired. There are two types of spectral criticality in existing work: spectral perturbation analysis based \cite{feng16,feng18,feng2019grass} and effective resistance based \cite{fegrass,kmp,EffRes08}.
	
	Suppose $S=(V,E_{S},w)$ is an initial subgraph. It is desired to calculate spectral criticality for each off-subgraph edge $(p,q) \in E \setminus E_{S}$. In spectral perturbation analysis, the dominant generalized eigenvector $h_t$ is first calculated using $t-$step power iterations:
	
	\begin{equation}
	\label{equ:pwriter}
	\setlength{\abovedisplayskip}{3pt}
	\setlength{\belowdisplayskip}{3pt}
	h_t=(L_S^{-1}L_G)^th_0 ~,
	\end{equation}
	where $h_0$ is a random vector. Then the Laplacian quadratic form is utilized to compute spectral criticality:
	\begin{equation}
	\setlength{\abovedisplayskip}{3pt}
	\setlength{\belowdisplayskip}{3pt}
	\label{equ:qform}
	h_t^T(L_G-L_S)h_t=\sum_{(p,q) \in E_G \setminus E_S} w_{p,q} (h_t^Te_{p,q})^2 ~,
	\end{equation} 
	where $e_{p,q}=e_p-e_q$ and $e_p$ is the $p$-th column of identity matrix. Smaller quadratic form indicates higher spectral similarity. To improve the spectral similarity, the edges with larger $w_{p,q} (h_t^Te_{p,q})^2$ should be recovered to the final subgraph.
	
	In effective resistance based method, $w_{p,q}R_S(p,q)$ reflects spectral importance of  off-subgraph edge $(p,q)$, where $R_S(p,q)$ denotes the effective resistance across nodes $p$ and $q$ in $S$ and satisfies:
	\begin{equation}
	\setlength{\abovedisplayskip}{3pt}
	\setlength{\belowdisplayskip}{3pt}
	\label{equ:effres}
	R_S(p,q)=e_{p,q}^TL_S^{-1}e_{p,q} ~.
	\end{equation}
	
	The main advantage of spectral perturbation analysis based method over effective resistance based method is that the former spectral criticality can be computed efficiently with respect to general graphs, while the latter can be computed efficiently only with respect to trees. So, the former can be easily combined with the iterative densification scheme proposed in \cite{feng18,feng2019grass}, which iteratively adds a small portion of off-tree edges and updates spectral criticality with respect to the current subgraph. The resulted graph sparsification algorithm GRASS \cite{feng2019grass} is more effective than effective resistance based algorithm feGRASS \cite{fegrass}, which means that GRASS can produce sparsifiers with same number of edges yet higher approximation level (lower relative condition number). 
	
	%
	
	\section{Spectral Graph Sparsification via Approximate Trace Reduction}
	\label{sec:method}
	
	In this section, we propose a novel graph spectral sparsification algorithm which is more effective than the state-of-the-art method.
 We first introduce the spectral criticality metric based on reducing the trace of $L_S^{-1}L_G$. Then, we propose a physics-inspired truncation strategy and an approach based on approximate inverse of Cholesky factor for computing approximate trace reduction efficiently. Finally, we present the effective graph sparsification algorithm. 
	
	\subsection{The Idea of Trace Reduction}
To obtain a high-quality spectral sparsifier $S$ for $G$, it is desirable to minimize the relative condition number $\kappa(L_G,L_S)$. We have\footnote{Recall that in practice $L_G$ and $L_S$ are the Laplacian matrices plus small positive diagonal elements. Therefore, the smallest generalized eigenvalue of $L_G$ and $L_S$ is $1$. It derives  $\kappa(L_G,L_S)=\lambda_{\max}(L_S^{-1}L_G)/\lambda_{\min}(L_S^{-1}L_G)=\lambda_{\max}(L_S^{-1}L_G)$.}:  
	\begin{equation}
	\setlength{\abovedisplayskip}{3pt}
	\setlength{\belowdisplayskip}{3pt}
	\kappa(L_G,L_S) = \lambda_{\max}(L_S^{-1}L_G) \le Trace(L_S^{-1}L_G) ~, 
	\end{equation}
where $Trace(L_S^{-1}L_G)$ is the trace of matrix $L_S^{-1}L_G$, i.e. the sum of matrix diagonal entries. We see that the trace of $L_S^{-1}L_G$ can be regarded as a proxy of the relative condiction number.
	 To improve the spectral similarity of $S$, we aim to reduce the  trace of $L_S^{-1}L_G$ as much as possible by recovering a few spectrally critical off-subgraph edges. This trace decreases as off-subgraph edges are recovered, and so does the relative condition number. Some off-subgraph edges lead to large trace reduction thus are spectrally critical, and should be first added into the sparsifier.  We now consider the effect of recovering an off-subgraph edge $(p,q)$ on the trace $Trace(L_S^{-1}L_G)$.
	
	After adding $(p,q)$ to $S$, the Laplacian matrix becomes:
	\begin{equation}
	\setlength{\abovedisplayskip}{3pt}
	\setlength{\belowdisplayskip}{3pt}
	L_{S^{'}}=L_S+w_{p,q}e_{p,q}e_{p,q}^T ~.
	\end{equation}
	Based on Sherman-Morrison Formula, we derive:
	\begin{equation}
	\setlength{\abovedisplayskip}{3pt}
	\setlength{\belowdisplayskip}{3pt}
	L_{S^{'}}^{-1}=L_S^{-1}-\frac{w_{p,q}L_S^{-1}e_{p,q}e_{p,q}^TL_S^{-1}}{1+w_{p,q}e_{p,q}^TL_S^{-1}e_{p,q}} ~.
	\end{equation}
	So the trace becomes:
	\begin{equation}
	\setlength{\abovedisplayskip}{3pt}
	\setlength{\belowdisplayskip}{3pt}
	\label{equ:equ1}
	Trace(L_{S^{'}}^{-1}L_G)=Trace(L_S^{-1}L_G)-\frac{w_{p,q}Trace(L_S^{-1}e_{p,q}e_{p,q}^TL_S^{-1}L_G)}{1+w_{p,q}e_{p,q}^TL_S^{-1}e_{p,q}}
	\end{equation}
	Due to $L_G=\sum_{(i,j) \in E} w_{i,j}e_{i,j}e_{i,j}^T$, we have:
	\begin{equation}
	\setlength{\abovedisplayskip}{3pt}
	\setlength{\belowdisplayskip}{3pt}
	\label{equ:equ2}
	Trace(L_S^{-1}e_{p,q}e_{p,q}^TL_S^{-1}L_G)=\sum_{(i,j) \in E} w_{i,j} (e_{i,j}^TL_S^{-1}e_{p,q})^2 ~.
	\end{equation} 
	Substituting (\ref{equ:equ2}) and (\ref{equ:effres}) into (\ref{equ:equ1}), we obtain
	\begin{equation}
	\setlength{\abovedisplayskip}{3pt}
	\setlength{\belowdisplayskip}{3pt}
	\label{equ:trred}
	Trace(L_{S^{'}}^{-1}L_G)=Trace(L_S^{-1}L_G)-\frac{w_{p,q}\sum_{(i,j) \in E} w_{i,j} (e_{i,j}^TL_S^{-1}e_{p,q})^2}{1+w_{p,q}R_S(p,q)} ~.
	\end{equation}
	We call the last term in (\ref{equ:trred}) as \textbf{ trace reduction} of off-subgraph edge $(p,q)$ with respect to subgraph $S$:
	\begin{equation}
	\setlength{\abovedisplayskip}{3pt}
	\setlength{\belowdisplayskip}{3pt}
	\label{equ:trreddef}
	TrRed_S(p,q)=\frac{w_{p,q}\sum_{(i,j) \in E} w_{i,j} (e_{i,j}^TL_S^{-1}e_{p,q})^2}{1+w_{p,q}R_S(p,q)} ~.
	\end{equation}
	
	Trace reduction reflects spectral importance of off-subgraph edges thus can be leveraged as a spectral criticality metric. However, using (\ref{equ:trreddef}) to compute spectral criticality for all off-subgraph edges leads to unacceptable $\Omega(m^2)$ complexity. Below we present an efficient physics-inspired truncation of trace reduction.
	In (\ref{equ:trreddef}), to compute spectral criticality of edge $(p,q)$, the summation is made over all edges. To reduce the complexity, it can be made only over edges $e=(i,j)$ with large $e_{i,j}^TL_S^{-1}e_{p,q}$. Note that the physical meaning of $e_{i,j}^TL_S^{-1}e_{p,q}$ is the voltage drop between $i$ and $j$ when an unit current flows into the subgraph $S$ at $p$ and leaves at $q$. Obviously, the nodes around $p$ have high electric potential and the nodes around $q$ have low electric potential. The edges between high-voltage and low-voltage nodes have large $e_{i,j}^TL_S^{-1}e_{p,q}$. Let $Nbr(p,\beta)$ denote the nodes found by $\beta$-layer breadth-first-search (BFS) from node $p$, then $Nbr(p,\beta)$ are high-voltage nodes and $Nbr(q,\beta)$ are low-voltage nodes. So the \textbf{truncated trace reduction} of off-subgraph edge $(p,q)$ with respect to subgraph $S$ can be computed with:
	\begin{equation}
	\setlength{\abovedisplayskip}{3pt}
	\setlength{\belowdisplayskip}{3pt}
	\label{equ:ttrreddef}
	tTrRed_S(p,q,\beta)=\frac{w_{p,q}}{1+w_{p,q}R_S(p,q)} \sum_{
		\mbox{\tiny$\begin{array}{c}
			(i,j) \in E\\
			i \in Nbr(p,\beta)\\
			j \in Nbr(q,\beta)\end{array}$}
	} w_{i,j} (e_{i,j}^TL_S^{-1}e_{p,q})^2 ~.
	\end{equation}
	

	\subsection{Approximately Computing Trace Reduction via Approximate Inverse of Cholesky Factor}
	
	Computing the truncated trace reduction (\ref{equ:ttrreddef}) can be too costly for large-scale problems because a linear equation for $L_S^{-1}e_{p,q}$ need to be solved for each edge $(p,q)$. We now show how to compute it efficiently. 
	If $S$ is a tree, we first run Tarjan's offline least common ancestor (LCA) algorithm \cite{lcacomp} to compute effective resistances for each off-subgraph edge $(p,q)$,  denoted as $R_S(p,q)$. 
	Consider the aforementioned physical model where an unit current flows into the subgraph $S$ through $p$ and leaves through $q$. Because $S$ is a tree, there exists a unique path from node $p$ to node $q$. We denote the path as $Path_S(p,q)$. The current only flows through $Path_S(p,q)$ so only the edges on that path result in voltage drop. We can assume the voltage of node $p$ equals to $R_S(p,q)$ and the voltage of node $q$ equals to $0$, which are denoted as $v(p)=R_S(p,q)$ and $v(q)=0$.
	
	The voltages of nodes $Nbr(p,\beta)$ can be computed  using BFS. Once a node $i$ is visited by BFS, we can first find its predecessor, which is denoted as $pred(i)$. Then check if the edge $(pred(i),i)$ lies on the unique path from $p$ to $q$. The voltage of node $i$ becomes:
	\begin{equation}
	\setlength{\abovedisplayskip}{3pt}
	\setlength{\belowdisplayskip}{3pt}
	v(i)\!=\!\left\{
	\begin{aligned}
	& v(pred(i))\!-\!\frac{1}{w_{pred(i),i}},~ \textrm{if} ~  (pred(i),i)\! \in \!Path_S(p,q) \\
	& v(pred(i)), ~ \textrm{otherwise} ~. \\
	\end{aligned}
	\right.
	\end{equation}
	It is the same with the nodes $Nbr(q,\beta)$ except that the voltage of newly discovered node $i$ is computed with:
	\begin{equation}
	\setlength{\abovedisplayskip}{3pt}
	\setlength{\belowdisplayskip}{3pt}
	v(i)\!=\!\left\{
	\begin{aligned}
	& v(pred(i))\!+\!\frac{1}{w_{pred(i),i}},~ \textrm{if} ~  (pred(i),i) \!\in\! Path_S(p,q) \\
	& v(pred(i)), ~ \textrm{otherwise} ~. \\
	\end{aligned}
	\right.
	\end{equation}
	
	After obtaining the voltages of $Nbr(p,\beta)$ and $Nbr(q,\beta)$, the truncated trace reduction of edge $(p,q)$ can be computed with:
	\begin{equation}
	\setlength{\abovedisplayskip}{3pt}
	\setlength{\belowdisplayskip}{3pt}
	\label{equ:ttrredtree}
	tTrRed_S(p,q,\beta)=\frac{w_{p,q}}{1+w_{p,q}R_S(p,q)} \sum_{
		\mbox{\tiny$\begin{array}{c}
			(i,j) \in E\\
			i \in Nbr(p,\beta)\\
			j \in Nbr(q,\beta)\end{array}$}
	} w_{i,j} (v(i)-v(j))^2 ~.
	\end{equation}
	
	The above method does not work for the case where $S$ is a general graph because there are possibly multiple paths from node $p$ to node $q$. All the edges on those paths result in voltage drop so the voltages of $Nbr(p,\beta)$ and $Nbr(q,\beta)$ cannot be computed using BFS. The main bottleneck in computing truncated trace reduction in (\ref{equ:ttrreddef}) is to compute $e_{i,j}^TL_S^{-1}e_{p,q}$. Suppose $L_S$ is factorized with Cholesky factorization: $L_S=LL^T$,
	where $L$ is a lower triangular matrix. Then we have:
	\begin{equation}
	\setlength{\abovedisplayskip}{3pt}
	\setlength{\belowdisplayskip}{3pt}
	\label{equ:fac}
	e_{i,j}^TL_S^{-1}e_{p,q}\!=\!(L^{-1}e_{i,j})^T(L^{-1}e_{p,q})\!=\!(L^{-1}e_i\!-\!L^{-1}e_j)^T(L^{-1}e_p\!-\!L^{-1}e_q) ~.
	\end{equation}
	$L^{-1}e_i$ is the $i$-th column of $L^{-1}$. If $L^{-1}$ is available, $e_{i,j}^TL_S^{-1}e_{p,q}$ can be computed efficiently with (\ref{equ:fac}), where only vector additions and inner product operations are required. However, computing and storing $L^{-1}$ explicitly is unacceptable for large-scale problems because $L^{-1}$ has much more nonzeros than $L$.
	
	To overcome this difficulty, we first observe that the majority of the entries in $L^{-1}$ are extremely small and discarding them does not cause large errors in computing $(L^{-1}e_i-L^{-1}e_j)^T(L^{-1}e_p-L^{-1}e_q)$. In fact, this is the key motivation for sparse approximate inverse (SPAI) techniques \cite{spai}. However, in general SPAI techniques, the structural information of $L$ is not taken into account. SPAI techniques are impractical for large-scale problems because it can be extremely time-consuming. In this work, we present a novel method for computing sparse approximations to inverse of Cholesky factor. The structural information of $L$ is utilized so the proposed method can be much more efficient than general SPAI techniques.
	
	Let $Z=L^{-1}=[z_1,z_2,..,z_n]$. Two useful properties of $L$ and $L^{-1}$ are listed below. 
	
	\begin{prop}
	\label{prop:prop1}
	    All the diagonal elements in $L$ are positive and all the off-diagonal elements in $L$ are nonpositive. $Z$ is lower triangular and all the elements in $Z$ are nonnegative.
	\end{prop}
	
	\begin{prop}
	\label{prop:prop2}
	    The columns of $Z$ satisfies:
	    \begin{equation}
	\setlength{\abovedisplayskip}{3pt}
	\setlength{\belowdisplayskip}{3pt}
	\label{equ:transition}
	z_j=\frac{1}{L_{j,j}}e_j+\sum_{i >j \& L_{i,j} \ne 0}\frac{-L_{i,j}}{L_{j,j}} z_i ~.
	\end{equation}
	\end{prop}

	Let $\tilde{z_i}$ denote the sparse approximation to $z_i$. Using Proposition \ref{prop:prop2}, $z_j$ can be computed approximately with:
	\begin{equation}
	\setlength{\abovedisplayskip}{3pt}
	\setlength{\belowdisplayskip}{3pt}
	z_j \approx z_j^* = \frac{1}{L_{j,j}}e_j+\sum_{i >j \& L_{i,j} \ne 0}\frac{-L_{i,j}}{L_{j,j}} \tilde{z_i} ~,
	\end{equation}
	which can be computed efficiently because $\tilde{z_i}s$ are sparse. 
	If $\Vert \tilde{z_i}-z_i \Vert \le \varepsilon$, $z_j^*$ approximates $z_j$ well:
	\begin{equation}
	\setlength{\abovedisplayskip}{3pt}
	\setlength{\belowdisplayskip}{3pt}
	\begin{aligned}
	\Vert z_j^*-z_j \Vert &= \Vert\sum_{i >j \& L_{i,j} \ne 0} \frac{-L_{i,j}}{L_{j,j}} (\tilde{z_i}-z_i)\Vert \!\le\! \sum_{i >j \& L_{i,j} \ne 0} \frac{-L_{i,j}}{L_{j,j}} \Vert\tilde{z_i}-z_i\Vert \\
	& \le \varepsilon \sum_{i >j \& L_{i,j} \ne 0} \frac{-L_{i,j}}{L_{j,j}} \le \varepsilon ~.\\
	\end{aligned}
	\end{equation}
	To maintain the sparsity, $z_j^*$ needs to be prunned. Proposition \ref{prop:prop1} implies that all the elements in ${z_j^*}$ are nonnegative so we just use a simple threshold based prunning strategy, where the elements smaller than the maximum element times a  threshold are set to $0$. The resulted  $\tilde{z_j}$ is sparse and  approximates $z_j$ well. The overall algorithm for computing sparse approximate inverse of Cholesky factor is described as Algorithm \ref{alg:iinv}.
	\begin{algorithm}[H]
		\caption{Sparse Approximate Inverse of the Cholesky Factor}
		\label{alg:iinv}
		\begin{algorithmic}[1]
			\setcounter{ALC@unique}{0}
			\REQUIRE Cholesky factor of $L_S$: $L$, a user-defined threshold $\delta$. 
			\ENSURE A sparse approximation to $L^{-1}$: $\tilde{Z}$.
			\FOR{$j=n$ to $1$}
			\STATE Compute $z_j^* = \frac{1}{L_{j,j}}e_j+\sum_{i >j \& L_{i,j} \ne 0}\frac{-L_{i,j}}{L_{j,j}} \tilde{z_i} ~$.
			\IF{$nnz(z_j^*) \le \log n $}
			\STATE $\tilde{z_j}=z_j^*$.
			\STATE Continue.
			\ENDIF
			\STATE Zero the elements in $z_j^*$ smaller than $\delta \max(z_j^*)$ to get $\tilde{z_j}$.	
			\ENDFOR
		\end{algorithmic}
	\end{algorithm}
	
	Using the sparse approximate inverse matrix $\tilde{Z}$, truncated trace reduction becomes ($\tilde{z}_{i,j}$ denotes $\tilde{z}_i-\tilde{z}_j$):
	\begin{equation}
	\setlength{\abovedisplayskip}{3pt}
	\setlength{\belowdisplayskip}{3pt}
	\label{equ:ttrrediinv}
	tTrRed_S(p,q,\beta) \!\approx\! \frac{w_{p,q}}{1\!+\!w_{p,q}\tilde{z}_{p,q}^T\tilde{z}_{p,q}} \!\sum_{
		\mbox{\tiny$\begin{array}{c}
			(i,j) \in E\\
			i \in Nbr(p,\beta)\\
			j \in Nbr(q,\beta)\end{array}$}
	} w_{i,j} (\tilde{z}_{i,j}^T\tilde{z}_{p,q})^2 ~,
	\end{equation}

	The time complexity of computing approximate trace reduction is closely related to the number of nonzeros in $\tilde{Z}$. In our experiments, the number of nonzeros in $\tilde{Z}$ is about $n \log n$ when the parameter $\delta$ in Alg. \ref{alg:iinv} is set to $0.1$. Here we just assume $nnz(\tilde{Z})=O(n \log n)$. If the parameter $\beta$ in (\ref{equ:ttrrediinv}) is small and fixed (e.g. $\beta=5$ in our experiments), the number of terms in the summation in (\ref{equ:ttrrediinv}) can be seen as a constant. So computing approximate trace reduction with (\ref{equ:ttrrediinv}) for one edge takes $O(\log n)$ time. Consequently, approximate trace reduction of all off-subgraph edges can be computed in $O(m \log n)$ time. Note that computing spectral criticality for all off-subgraph edges in GRASS \cite{feng2019grass} also takes $O(m \log n)$ time, so the time complexity of the proposed method is 
same as that of GRASS.
	
	\subsection{The Overall Algorithm}
	
	To develop a more effective graph sparsification algorithm, we combine the proposed approximate trace reduction based approach with the iterative densification scheme  in \cite{feng18} and the strategy for excluding similar off-subgraph edges in \cite{fegrass}. The overall flow of the proposed graph sparsification algorithm is described as Algorithm \ref{alg:main}. In Step 1, the low-stretch spanning tree can be constructed with the approach of maximum effective weight spanning tree (MEWST) proposed in \cite{fegrass}, or other efficient approaches.
	
	\begin{algorithm}[h]
		\caption{Graph Spectral Sparsification via Approximate Trace Reduction}
		\label{alg:main}
		\begin{algorithmic}[1]
			\setcounter{ALC@unique}{0}
			\REQUIRE Graph $G=(V,E,w)$, the desired number of edges to recover $\alpha$, the number of iterations for recovering edges $N_r$.
			\ENSURE Sparsifier $P$. 
			\STATE Extract a low-stretch spanning tree $T$ from $G$. Set $P=T$.
			\STATE Compute truncated trace reduction for off-tree edges with (\ref{equ:ttrredtree}). Sort off-tree edges by truncated trace reduction from the largest to the smallest, to get an edge list \emph{OffTreeEdges}.
			\STATE $\emph{count}=0,~k=1$.
			\WHILE{$\emph{count} \le \frac{\alpha}{N_r}$}
			\STATE Get edge $(i,j)=$\emph{OffTreeEgdes}$[k]$, $k$++. 
			\IF{$(i,j)$ is not marked}
			\STATE Add $(i,j)$ into $P$. $\emph{count}$++.
			\STATE Mark the edges similar to $(i,j)$ for exclusion from recoverage using the technique proposed in \cite{fegrass}.
			\ENDIF
			\ENDWHILE
			\FOR{$iter=2$ to $N_r$}
			\STATE Factorize Laplacian matrix of the latest subgraph $L_P$.
			\STATE Run Alg. \ref{alg:iinv} to obtain the approximate inverse matrix $\tilde{Z}$.
			\STATE Compute approximate trace reduction for off-subgraph edges with (\ref{equ:ttrrediinv}). Sort them by approximate trace reduction from the largest to the smallest to get an edge list \emph{OffSubgraphEdges}.
			\STATE $\emph{count}=0,~k=1$.
			\WHILE{$\emph{count} \le \frac{\alpha}{N_r}$}
			\STATE Get edge $(i,j)=$\emph{OffSubgraphEgdes}$[k]$, $k$++.
			\IF{$(i,j)$ is not marked}
			\STATE Add $(i,j)$ into $P$. $\emph{count}$++.
			\STATE Mark the edges similar to $(i,j)$ for exclusion from recoverage using the technique proposed in \cite{fegrass}.
			\ENDIF
			\ENDWHILE
			\ENDFOR
		\end{algorithmic}
	\end{algorithm} 
	
	\section{Experimental Results}
	We have implemented the proposed  algorithm (Alg. 2) for graph sparsification, including the MEWST algorithm in \cite{fegrass} to construct a low-stretch spanning tree. A PCG solver is also implemented, which takes the output of Alg. 2 and GRASS \cite{feng2019grass} and then factorizes the sparsifier's Laplacian matrix   with CHOLMOD \cite{cholmod2008} before performing PCG iteration. The programs are written in C++. The result of GRASS is obtained by running the GRASS program \cite{grassweb}. All experiments are carried out using a single CPU core of a computer with Intel Xeon E5-2630 CPU @2.40 GHz and 256 GB RAM.


	\subsection{Results for Graph Spectral Sparsification}
	
	In this subsection, we compare the proposed Alg. 2 with the state-of-the-art method GRASS \cite{feng2019grass}. The test cases are the undirected graphs used in \cite{feng2019grass}, supplemented by five larger cases derived from 2D finite-element triangular meshes. They are all available from  SuiteSparse Matrix Collection \cite{sparse}. The results are listed in Table 1. For both our Alg. 2 and GRASS, the sparsifier is constructed by recovering $10\%|V|$ off-tree edges, and a five-iteration edge-recovering strategy is adopted (i.e. recovering $2\%|V|$ off-subgraph edges  in each iteration). $T_s$, $\kappa$, $N_i$ and $T_i$ are the time for constructing the sparisifier $P$, the relative condition number $\kappa(L_G,L_P)$,  the number of iteration steps and the time for PCG iteration, respectively.  The relative tolerance for PCG convergence is set to $10^{-3}$ and the right-hand-side (RHS) vector is generated randomly.
	\begin{table}[h]
		\centering
 		\setlength{\abovecaptionskip}{3pt}
		\caption{Results for spectral graph sparsification (time in unit of second, $\kappa$ means the relative condition number)}
		\label{table2}
		\small{
			\begin{tabular}{@{~}c@{~}c@{~}c@{~}c@{~}c@{~}c@{~}c@{~}c@{~}c@{~}c@{~}c@{~}c@{~}c@{~}c@{~}c@{~}c@{~}c@{~}c@{~}c@{~}c@{~}c@{~}c@{~}c@{~}c@{~}c@{~}c@{~}c@{~}c@{~}c@{~}}
				\hline
				\multirow{2}{*}{Case} & \multirow{2}{*}{|V|} & \multirow{2}{*}{|E|} & & \multicolumn{4}{c}{GRASS} & &  \multicolumn{4}{c}{Proposed} & & \multicolumn{2}{@{}c@{~}}{Reduction} \\
				\cline{5-8} \cline{10-13} \cline{15-16} 
				& & &  & $T_s$ & $\kappa$ & $N_i$ & $T_i$ & & $T_s$ & $\kappa$ & $N_i$ & $T_i$ & & $\kappa$ & $T_i$ \\
				\hline
				ecology2 & 1.0E6 & 2.0E6 & & 11.0 & 108 & 43 & 1.98 & & 11.8 & 41.3 & 28 & 1.20 & & 2.6X & 1.7X\\ 
				\hline
				thermal2 & 1.2E6 & 3.7E6 & & 15.1 & 85.5 & 63 & 4.25 & & 19.3 & 45.1 & 46 & 2.71 & & 1.9X & 1.6X\\
				\hline 
				parabolic & 0.5E6 & 1.6E6 & & 4.62 & 205 & 57 & 1.26 & & 6.35 & 42.8 & 29 & 0.61 & & 4.8X & 2.1X\\
				\hline
				tmt\_sym & 0.7E6 & 2.2E6 & & 7.72 & 149 & 56 & 1.88 & & 9.86 & 43.0 & 29 & 0.96 & & 3.5X & 2.0X\\
				\hline
				G3\_circuit & 1.6E6 & 3.0E6 & & 16.5 & 72.7 & 56 & 4.89 & & 13.5 & 63.8 & 50 & 3.45 & & 1.1X & 1.1X\\
				\hline
				NACA0015 & 1.0E6 & 3.1E6 & & 16.6 & 137 & 84 & 4.66 & & 22.1 & 54.5 & 52 & 2.38 & & 2.5X & 2.0X\\
				\hline
				M6 & 3.5E6 & 1.1E7 & & 80.3 & 185 & 95 & 17.0 & & 94.0 & 60.8 & 58 & 9.47 & & 3.0X &1.8X \\
				\hline
				333SP & 3.7E6 & 1.1E7 & & 59.0 & 166 & 94 & 16.9 & & 86.7 & 89.2 & 69 & 11.3 & & 1.9X&1.5X\\
				\hline
				AS365 & 3.8E6 & 1.1E7 & & 87.2 & 145 & 95 & 18.8 & & 100 & 55.8 & 56 & 10.0 & & 2.6X & 1.9X\\
				\hline
				NLR & 4.2E6 & 1.2E7 & & 98.1 & 142 & 94 & 20.4 & & 114 & 73.2 & 60 & 11.8 & & 1.9X & 1.7X\\
				\hline
				Average & - & - & & - & - & - & - & & - & - & - & - & & 2.6X & 1.7X\\
				\hline 
			\end{tabular}	
		}
	\end{table}
	
	From the table we see that the time for graph sparsification ($T_s$) is comparable between GRASS and the proposed algorithm, but the latter produces sparisifiers with better quality. The sparsifier generated with the proposed algorithm makes the relative condition number 2.6X smaller than that derived by the sparsifier generated with GRASS, on average.
	With the sparsifier's Laplaician matrix as the preconditioner, the PCG iteration consumes less time, which is averagely 1.7X smaller than that caused by the preconditioner produced by GRASS (i.e. a 41\% reduction).

	\subsection{Results for Power Grid Transient Simulation}
	In this subsection, we apply the graph spectral sparsification to power grid (PG) transient simulation. Test cases are from two well-known power grid benchmarks \cite{ibmpg,thupg}. Interconnect capacitance plays an increasingly important role in circuit simulation \cite{yu2012efficient} and is included in IBM PG benchmarks. 	
	For the cases from \cite{thupg}, capacitances with values randomly ranging from 1pF to 10pF are added (similar to IBM PG benchmarks) and periodic pulse currents are generated at each current source for transient analysis.
	
	Transient analysis of power grid can be formulated as differential algebra equations (DAEs) via modified nodal analysis. With time integration schemes like backward Euler scheme or trapezoidal scheme, the DAEs are converted to a set of linear equation systems for solving the node voltages at consecutive time points. In our experiment, we use the backward Euler scheme and solve the following linear equation at each time point:
	\begin{equation}
	\setlength{\abovedisplayskip}{3pt}
	\setlength{\belowdisplayskip}{3pt}
	\label{eq:trans}
	(L_G+\frac{C}{h}) x(t+h)=\frac{C}{h} x(t)+u(t) ~,
	\end{equation}
	where $C$ is the matrix for capacitive and inductive elements,
	$L_G$ denoting the Laplacian matrix of PG graph $G$ is the matrix for conductance and resistance, $u(t)$ is the vector of current sources. $x(t)$ and $x(t+h)$ stand for the vectors of node voltages at two consecutive time points with interval (time step) $h$. Given an initial condition $x(0)$ obtained from DC analysis, the transient simulation towards time 5ns is performed via repeatedly solving (\ref{eq:trans}).

	Direct sparse matrix solver combined with a strategy of fixed time step can be very efficient for the PG transient simulation \cite{powerrushtrans}. It only requires one matrix factorization at the beginning of the transient simulation. Then, with fixed time step $h$, the following transient computation requires only forward/backward substitutions. However, the maximum step size is limited by the smallest distance among the breakpoints of current source waveforms. If varied time steps are adopted, the direct solver can be extremely time-consuming due to the expensive matrix factorizations performed whenever the time step changes. 
	
	The iterative solver is more suitable for varied time steps  and can be accelerated by the spectral graph sparsifiers. This allows larger time steps to reduce the total time for solving (\ref{eq:trans}) in transient simulation, so as to make the iterative solver competitive to the direct solver in this application.
	In our experiments, the PCG solvers using GRASS and proposed algorithm for constructing preconditioner are tested. The sparsifier is extracted by recovering $10\%|V|$ off-tree edges and the preconditioner constructed in DC analysis is used for all the following transient steps. The relative tolerance of PCG convergence is set to $10^{-6}$. The varied time steps are determined with the breakpoints of current sources, but restricted not to exceed 200ps for error control. The computational costs are listed in Table 2, where $T_\textrm{tr}$ and $T_s$ denote the runtime for transient simulation and graph sparsification respectively, $N_a$ denotes the average number of PCG iterations for solving one time step, Mem denotes the memory usage. The memory usage of the GRASS derived iterative solver is not listed as it always equals to that of ours. From the results we see that they both cost similar time for sparsification, and the proposed algorithm leads to 1.4X less time for transient simulation, averagely (see column of Sp$_2$). It means the proposed algorithm brings about 30\% reduction of time for PG transient simulation compared with the iterative solver using GRASS.
	\begin{table}[h]
		\centering
 		\setlength{\abovecaptionskip}{3pt}
		\caption{Results for power grid transient simulation (time in unit of second, $N_a$ means average iteration number)}
		\label{table3}
		\small{
			\begin{tabular}{@{~}c@{~}c@{~}c@{~}c@{~}c@{~}c@{~}c@{~}c@{~}c@{~}c@{~}c@{~}c@{~}c@{~}c@{~}c@{~}c@{~}c@{~}}
				\hline
				\multirow{2}{*}{Case}& \multirow{2}{*}{|V|} & & \multicolumn{2}{c}{Direct} & & \multicolumn{3}{c}{GRASS} & & \multicolumn{4}{c}{Proposed} & & \multicolumn{2}{@{}c@{}}{Speedup}\\
				\cline{4-5} \cline{7-9} \cline{11-14} \cline{16-17}
				& &  & $T_\textrm{tr}$ & Mem & & $T_s$ & $T_\textrm{tr}$ & $N_a$ & & $T_s$ & $T_\textrm{tr}$ & $N_a$ & Mem & & Sp$_1$ & Sp$_2$\\
				\hline
				ibmpg3t & 8.5E5 & &111 & 0.9GB & & 6.79 & 52.9 & 17.0 & & 6.85 & \textbf{35.9} & 11.1  & 0.2GB && 3.1 & 1.5 \\
				\hline
				ibmpg4t & 9.5E5 & &158 & 1.1GB & & 11.0 & 39.7 & 21.5 & & 11.2 & \textbf{32.9} & 18.1 & 0.3GB && 4.8 & 1.2\\
				\hline
				ibmpg5t & 1.1E6 & &74.8 & 0.7GB & & 9.57 & 85.1 & 22.1 & & 9.31 & \textbf{65.7} & 17.8 & 0.3GB & & 1.1 & 1.3\\
				\hline
				ibmpg6t & 1.7E6 & &102 & 1.1GB & & 16.1 & 137 & 22.5 & & 17.4 & \textbf{97.0} & 15.3 & 0.5GB && 1.0 & 1.4\\
				\hline
				thupg1t & 5.0E6 & &698 & 4.2GB & & 76.0 & 176 & 20.8 & & 80.1 & \textbf{134} & 14.7 &  1.1GB && 5.2 & 1.3\\
				\hline
				thupg2t & 9.0E6 & &1203 & 7.4GB & & 167 & 330 & 20.7 & & 168 & \textbf{244} & 14.8 &  1.9GB && 4.9 & 1.4\\
				\hline
				Average & -&&-&-&&-&-&-&&-&-&-&-&& 3.4 & 1.4 \\
				\hline
			\end{tabular}	
		}
	\end{table}
	
\begin{figure}[h]
		\setlength{\abovecaptionskip}{3pt}
		\setlength{\belowcaptionskip}{0pt}
		\centering
		\includegraphics[width=7.5cm]{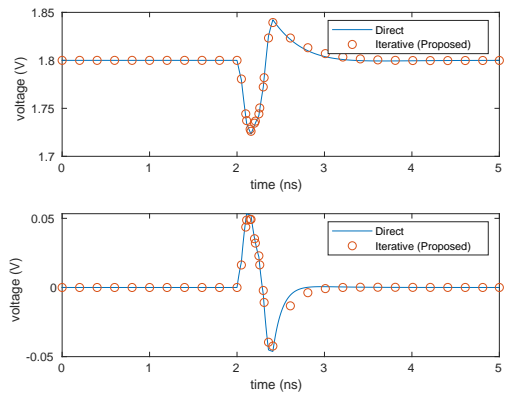}
		\caption{The transient simulation results of a VDD node (up) and a GND node (down) in case ``ibmpg4t'', obtained with direct equation solver and the proposed iterative solver.}
		\label{fig:fig2}
	\end{figure}

	
	To validate the effectiveness of using iterative solver in transient simulation, we also list the results derived from using direct solver \cite{cholmod2008} in Table 2. For the test cases, the fixed time step is set to 10ps due to the limit of the smallest distance among the current-source breakpoints. From Table 2 we see that the iterative solver leveraging the proposed graph sparsification runs averagely 3.4X faster than the direct solver for performing transient simulation (see Sp$_1$). On memory usage, the iterative solver also exhibits about 4X reduction. This advantage on memory cost should be more remarkable for larger test cases. We have also compared the node voltages computed with our iterative solver and the direct solver, whose results show that their difference is less than 16mV for all cases.  The transient waveforms of node n1\_7880\_8843 and node n0\_13206\_959 in case ``ibmpg4t'' are plotted in Fig. \ref{fig:fig2}. They validate the accuracy of  transient simulation using the proposed iterative solver.
	
\begin{figure}[h]  
		\setlength{\abovecaptionskip}{3pt}
		\setlength{\belowcaptionskip}{0pt}
		\centering
		\includegraphics[width=8.3cm]{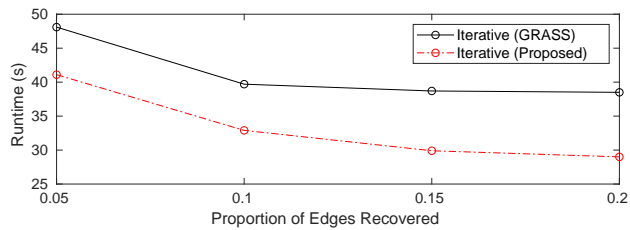}
		\caption{Tradeoff between the sparsity of sparsifier and runtime for power grid transient analysis, for  case ``ibmpg4t''.}
		\label{fig:fig1}
	\end{figure}

	
	To show the effect of sparsifier's sparsity on the runtime for transient analysis, we gradually increase the recovered off-tree edges for the case ``ibmpg4t'' and record the resulted transient simulation runtime. They are plotted in Fig. \ref{fig:fig1}. Although adding more than $10\%|V|$ off-tree edges  leads to slightly shorter runtime, it  also results in larger memory usage. The figure also shows that with more off-tree edges recovered, the proposed algorithm exhibits larger advantage over GRASS on reducing the time for PG simulation.

	\subsection{Results for Spectral Graph Partitioning}
	
	In this subsection, the graph sparsifiers are leveraged to accelerate the computation of Fiedler vector, which is a key subroutine of spectral graph partitioning \cite{spectralpartition}. Fiedler vector is the eigenvector corresponding to the smallest nonzero eigenvalue of the graph Laplacian matrix, which is usually computed with the inverse power iteration. In each iteration step,
	the equation with graph Laplacian matrix is solved. For this problem, we compare the efficiency of direct solver \cite{cholmod2008} and two graph sparsification based PCG solvers. The test cases are from \cite{feng2019grass} and 5 steps of inverse power iteration are executed. The results are listed in Table 3, where $T_D$ ($T_I$) denotes the runtime of direct (iterative) solver, which includes the time for matrix factorization and inverse power iteration. $N_a$ denotes the average PCG iteration number in each step of inverse power iteration, 
	while RelErr denotes the ratio of nodes which are assigned to different partitions from the results of direct method.
	
	From Table 3 we can see, compared with direct solver, the graph sparsification based iterative solvers show advantages in both runtime (see Sp$_1$) and memory usage. And, the error caused by iterative solver is marginal. The proposed algorithm leads to 1.4X speedup averagely over the GRASS based solver (see Sp$_2$). This again validates the effectiveness of the proposed algorithm.
	
	\begin{table}[h]
		\centering
 		\setlength{\abovecaptionskip}{3pt}
		\caption{Results for computing approximate Fiedler vector (time in unit of second, $N_a$ means average iteration number)}
		\small{
			\begin{tabular}{@{~}c@{~}c@{~}c@{~}c@{~}c@{~}c@{~}c@{~}c@{~}c@{~}c@{~}c@{~}c@{~}c@{~}c@{~}c@{~}c@{~}c@{~}c@{~}}
				\hline
				\multirow{2}{*}{Case} & & \multicolumn{2}{c}{Direct} & & \multicolumn{3}{c}{GRASS} & & \multicolumn{4}{c}{Proposed} & & \multicolumn{2}{@{}c@{~}}{Speedup}\\
				\cline{3-4} \cline{6-8} \cline{10-13} \cline{15-16}
				& & $T_D$ & Mem & & $T_I$ & $N_a$ & RelErr & & $T_I$ & Mem & $N_a$ & RelErr & & Sp$_1$ & Sp$_2$ \\
				\hline 
				ecology2 & & 13.8 & 0.7GB & & 7.96 & 30.8 & 1.8E-3 & & \textbf{5.92} & 0.2GB & 20.2 & 2.0E-3 & & 2.3 & 1.3 \\
				\hline
				thermal2 & & 18.3 & 1.0GB & & 7.92 &  22.6 & 5.5E-3 & & \textbf{6.50} & 0.3GB & 16.8 & 4.8E-3 & & 2.8 & 1.2 \\
				\hline
				parabolic & & 5.17 & 0.4GB & & 2.60 &  20.0 & 5.6E-3 & & \textbf{1.86} & 0.1GB & 12.6 & 4.7E-3 & & 2.8 & 1.4 \\
				\hline
				tmt\_sym & & 14.0 & 0.5GB & & 5.83 &  32.0 & 3.9E-3 & & \textbf{3.74} & 0.1GB & 17.0 & 3.5E-3 & & 3.7 & 1.6\\
				\hline
				G3\_circuit & & 41.8 & 1.3GB & & 10.8 &  25.0 & 2.1E-4 & & \textbf{8.61} & 0.3GB & 22.0 & 2.3E-4 & & 4.9 & 1.3 \\
				\hline
				Average & &-&-&&-&-&-&&-&-&-&-&& 3.3 & 1.4 \\
				\hline
			\end{tabular}	
		}
	\end{table}
	
	\section{Conclusions}
	This paper presents a graph sparsification algorithm more effective than the state-of-the-art GRASS \cite{feng2019grass}. Based on the fact that the trace of  matrix $L_S^{-1}L_G$ can be regarded as a proxy of relative condition number of Laplacians $L_G$ and $L_S$, efficient techniques are developed to approximately compute the trace reduction caused by recovering each off-subgraph edge so as to identify the spectrally important edges. Combined with other advanced techniques, they  derive a highly effective graph sparsification algorithm. Experiments on   power grid transient simulation and other problems reveal the efficiency and robustness of the proposed methods.
	
	

	\bibliographystyle{ACM-Reference-Format}
	\bibliography{ref1}

\end{document}